\documentclass{nature}
\usepackage{aas_macros,graphicx}

\title{The unpolarized macronova associated with the gravitational wave event GW\,170817}

\author{S. Covino$^{\ref{INAFBr}}$, K. Wiersema$^{\ref{UKLei}}$, Y. Z. Fan$^{\ref{PMCh}}$, K. Toma$^{\ref{TOMAInst},\ref{TOMAInst2}}$, A. B. Higgins$^{\ref{UKLei}}$, A. Melandri$^{\ref{INAFBr}}$, P. D'Avanzo$^{\ref{INAFBr}}$, C. G. Mundell$^{\ref{BathUK}}$, E. Palazzi$^{\ref{INAFBO}}$, N. R. Tanvir$^{\ref{UKLei}}$, M. G. Bernardini$^{\ref{MGFr},\ref{INAFBr}}$, M. Branchesi$^{\ref{MarInst},\ref{MarInst2}}$, E. Brocato$^{\ref{BrocInst}}$, S. Campana$^{\ref{INAFBr}}$, S. di Serego Alighieri$^{\ref{INAFFi}}$, D. G\"otz$^{\ref{Gotz}}$, J. P. U. Fynbo$^{\ref{DARK}}$, W. Gao$^{\ref{INAFBr},\ref{GaoInst}}$, A. Gomboc$^{\ref{ANDREJAinst}}$, B. Gompertz$^{\ref{ST}}$, J. Greiner$^{\ref{GrInst}}$, J. Hjorth$^{\ref{DARK}}$, Z. P. Jin$^{\ref{PMCh}}$, L. Kaper$^{\ref{RaNL}}$, S. Klose$^{\ref{KloInst}}$, S. Kobayashi$^{\ref{LiverUK}}$, D. Kopac$^{\ref{Sloinst}}$, C. Kouveliotou$^{\ref{KouI}}$, A. J. Levan$^{\ref{Warw}}$, J. Mao$^{\ref{MAOCh}}$, D. Malesani$^{\ref{DARK}}$, E. Pian$^{\ref{INAFBO}}$, A. Rossi$^{\ref{INAFBO}}$, R. Salvaterra$^{\ref{Ruben}}$, R. L. C. Starling$^{\ref{UKLei}}$, I. Steele$^{\ref{LiverUK}}$, G. Tagliaferri$^{\ref{INAFBr}}$, E. Troja$^{\ref{EleUSA}}$, A. J. van der Horst$^{\ref{vanh},\ref{vanh2}}$ \& R. A. M. J. Wijers$^{\ref{RaNL}}$}

\begin{document}

\maketitle

\begin{affiliations}
 \item 
\label{INAFBr}
INAF / Brera Astronomical Observatory, via Bianchi 46, 23807, Merate (LC), Italy 
\item 
\label{UKLei}
Department of Physics and Astronomy, University of Leicester, Leicester LE1 7RH, UK
\item 
\label{PMCh} 
Key Laboratory of Dark Matter and Space Astronomy, Purple Mountain Observatory, Chinese Academy of Sciences, Nanjing 210008, China
\item
\label{TOMAInst}
Frontier Research Institute for
Interdisciplinary Sciences, Tohoku University, Sendai 980-8578, Japan
\item
\label{TOMAInst2}
Astronomical Institute, Tohoku University, Sendai 980-8578, Japan
\item
\label{BathUK}
Department of Physics, University of Bath, Claverton Down, Bath, BA2 7AY, UK
\item
\label{INAFBO}
INAF / Istituto di Astrofisica Spaziale e Fisica Cosmica di Bologna, Via Gobetti 101, 40129 Bologna, Italy
\item
\label{MGFr}
Laboratoire Univers et Particules de Montpellier, Université de Montpellier, CNRS/IN2P3, Montpellier, France 
\item
\label{MarInst}
Gran Sasso Science Institute (GSSI), 67100 L'Aquila, Italy 
\item
\label{MarInst2}
INFN / Laboratori Nazionali del Gran Sasso, 67100 L'Aquila, Italy
\item 
\label{BrocInst}
INAF / Osservatorio Astronomico di Roma, Via di Frascati, 33, 00078 Monteporzio Catone, Italy
\item 
\label{INAFFi}
INAF / Osservatorio Astrofisico di Arcetri, Largo E. Fermi 5, 50125, Firenze, Italy
\item
\label{Gotz}
CEA Saclay - DRF/Irfu/D\`epartement d'Astrophysique, 91191 Gif-sur-Yvette, France
\item
\label{DARK}
Dark Cosmology Centre, Niels Bohr Institute, University of Copenhagen, Juliane Maries Vej 30, 2100 Copenhagen, Denmark
\item
\label{GaoInst}
Department of Physics and Institute of Theoretical Physics, Nanjing Normal University, Nanjing 210046, China
\item
\label{ANDREJAinst}
Centre for Astrophysics and Cosmology, University of Nova Gorica, Vipavska 11c, 5270 Ajdov\v s\v cina, Slovenia
\item
\label{ST}
Space Telescope Science Institute, Baltimore, MD 21218, USA
\item
\label{GrInst}
Max-Planck-Institut f\"ur extraterrestrische Physik, Giessenbachstr. 1, 85748 Garching, Germany
\item
\label{RaNL}
Anton Pannekoek Institute, University of Amsterdam, Science Park 904, 1098XH Amsterdam
\item
\label{KloInst}
Th\"uringer Landessternwarte Tautenburg, Sternwarte 5, 07778 Tautenburg, Germany
\item
\label{LiverUK}
Astrophysics Research Institute, Liverpool John Moores University, ic2, Liverpool Science Park, 146 Brownlow Hill, Liverpool L3 5RF, UK
\item
\label{Sloinst}
Faculty  of  Mathematics  and  Physics,  University  of  Ljubljana, Jadranska 19, 1000 Ljubljana, Slovenia
\item
\label{KouI}
Department of Physics, The George Washington University, 725 21st Street NW, Washington, DC 20052, USA
\item
\label{Warw}
Department of Physics, University of Warwick, Coventry, CV4 7AL, UK
\item
\label{MAOCh}
Yunnan Observatories, Chinese Academy of Sciences, 650011 Kunming, Yunnan Province, China 
\item
\label{Ruben}
INAF / Istituto di Astrofisica Spaziale e Fisica Cosmica Milano, via E. Bassini 15, 20133, Milano, Italy
\item
\label{EleUSA}
Department of Astronomy, University of Maryland, College Park, MD 20742, USA 
\item
\label{vanh}
Department of Physics, The George Washington University, 725 21st Street NW, Washington, DC 20052, USA
\item
\label{vanh2}
Astronomy, Physics, and Statistics Institute of Sciences (APSIS)
\end{affiliations}

\begin{abstract}
The merger of two dense stellar remnants including at least one neutron star (NS) is predicted to produce gravitational waves (GWs) and short duration gamma ray bursts (GRBs)\cite{1989Natur.340..126E,2014ARA&A..52...43B}. In the process, neutron-rich material is ejected from the system and heavy elements are synthesized by r-process nucleosynthesis\cite{1989Natur.340..126E,1999A&A...341..499R}. 
The radioactive decay of these heavy elements produces additional transient radiation termed ``kilonova" or ``macronova"\cite{1998ApJ...507L..59L,2005ApJ...634.1202R,2010MNRAS.406.2650M,2013ApJ...774...25K,2013ApJ...775...18B,2013ApJ...775..113T,2017RPPh...80i6901B}. 
We report the detection of linear optical polarization $P = (0.50 \pm 0.07)\%$ at 1.46 days after detection of the GWs from GW\,170817, a double neutron star merger associated with an optical macronova counterpart and a short GRB\cite{gcn21505,gcn21507,gcn21509,gcn21529}. The optical emission from a macronova is expected to be characterized by a blue, rapidly decaying, component and a red, more slowly evolving, component due to material rich of heavy elements, the lanthanides\cite{2015MNRAS.450.1777K}. The polarization measurement was made when the macronova was still in its blue phase, during which there is an important contribution from a lanthanide-free outflow. The low degree of polarization is consistent with intrinsically unpolarized emission scattered by Galactic dust, suggesting a symmetric geometry of the emitting region and low inclination of the merger system. Stringent upper limits to the polarization degree from 2.45 - 9.48 days post-burst are consistent with the lanthanides-rich macronova interpretation.
\end{abstract}

The search for the optical counterpart to GW\,170817\cite{gcn21505} quickly allowed the discovery of the possible counterpart, named AT2017gfo (also known as SSS17a), in the outskirts of the elliptical galaxy NGC4993 at about 40\,Mpc\cite{gcn21529}. Spectroscopic observations showed that this source was highly unusual and likely associated with the GW event and a short GRB\cite{Pian}. Only a few possible detections of macronova emission have been reported in the literature, all by analyzing the temporal and spectral evolution of the afterglows of on-axis short GRBs\cite{2013Natur.500..547T,2015NatCo...6E7323Y,2015ApJ...811L..22J,2016NatCo...712898J}. GRB\,170817A is intrinsically the weakest short GRB detected so far, which may be a regular short GRB but viewed from an off-axis orientation. The off-axis scenario is also helpful in reconciling the probability of GW/GRB association for this event\cite{2016A&A...594A..84G,2017arXiv170807008J}. Therefore SSS17a  offers the unique opportunity to study a macronova emission plausibly not polluted by the GRB emission. The combination of a likely off-axis jet and the potential ability of LIGO/Virgo data to constrain the merger geometry and orientation, gave great urgency to a polarimetric measurement of the symmetry and orientation of the optical emission region(s) post-merger. Measuring the degree of polarization of the electromagnetic emission provides unique constraints on the geometry of the system and the properties of any incipient magnetic fields\cite{2013Natur.504..119M,2014Natur.509..201W}.

Our linear polarimetry campaign consisted of a set of five observations carried out with the European Southern Observatory/Very Large Telescope (ESO/VLT) equipped with the Focal Reducer/low dispersion Spectrograph (FORS2)\footnote{http://www.eso.org/sci/facilities/paranal/instruments/fors.html} starting on 18 August 2017 and spanning almost 10 days. After that, the transient was too faint for a reliable polarimetric analysis. Details about the observing setup, data reduction and analysis are reported in the ``Methods" section. The derived degree of linear polarization, position angle and optical brightness, after instrumental corrections have been applied, are given in Table\,\ref{tab:pol} (a complete observation log including date of the observations, exposure times, filters and seeing is reported in Table\,\ref{tab:obslog}). The Stokes parameters for optical transient and nearby field stars for the first four epochs are shown in Figure\,1. Over the duration of our campaign, the transient showed a degree of linear polarization and position angle fully consistent with that shown by stars in the field, whose polarization is induced by dust in our Galaxy. This implies that the macronova emission is essentially unpolarized at a level driven by the photometric uncertainties and the spread of polarization shown by field stars, i.e. 0.4-0.5\%.

GW\,170817 originated in the coalescence of two NSs\cite{gcn21505}. Numerical simulations show that these events can eject a small part of the original system into the interstellar medium\cite{1999A&A...341..499R,2011ApJ...738L..32G,2013PhRvD..88d4026H} and also form a centrifugally supported disk that is quickly dispersed in space with a neutron-rich wind\cite{2013ApJ...774...25K}. These two different ejection mechanisms are characterized by material of differing composition. The outflows from the disk are likely lanthanide-free since the synthesis of heavier elements is suppressed by the high temperature\cite{2013ApJ...775...18B}, while the surface material is the site of an intense r-process nucleosythesis, producing heavy elements. In both cases the spectrum should be close to a black-body, peaking in the optical in the disk-wind case and in the infrared for the lanthanide-rich material due to its very large opacity\cite{2013ApJ...775...18B,2013ApJ...775..113T}.
Ejecta should flow out anisotropically around the orbital plane of the system and outflows in the polar region can be produced by a strong shock driven by the merger and by processes such as viscous heating and magnetic effects in the disk\cite{2015PhRvD..92l4034K}. Anisotropies induced by electron scattering can then produce some polarization\cite{2015PhRvD..92d4028K}. As pointed out by\cite{2015PhRvD..92l4034K}, in case of high optical depth to electron scattering ($\sim 1$) and assuming spectral lines do not significantly depolarize the global emission, the linear polarization observed from the equatorial plane could be as high as a few percent. It depends, as is the case of supernovae (SNe), on the degree of asymmetry of the photosphere. However, with respect to the SN case, if the ejecta are mainly composed by r-process elements the ionization degree will not be particularly high\cite{2013ApJ...774...25K,2014ApJ...780...31T} and the number density of free electrons proportionally lower. With typical parameters the electron scattering opacity will be lower than the total opacity ($\kappa \sim 10^2$\,cm$^2$\,g$^{-1}$) by three orders of magnitudes, and then the expected linear polarization is reduced than in the electron-scattering dominant case by a similar factor\cite{2015PhRvD..92d4028K}.

The emission from the wind could instead have a different composition and be more similar to typical SN ejecta. A blue component was identified in the spectra of AT2017gfo\cite{Pian}, and it showed a more rapid evolution than the redder component. The emission from the lanthanide-rich material is not expected to display any detectable linear polarization, but the situation can be significantly different for the early emission phase that is dominated by the lanthanide-free material, for which the temperature is higher (so is the ionization degree) and the electron scattering optical depth is $\sim 1$. Assuming that at our first epoch the blue component was at least as important as the lanthanide-rich emission, we can derive an upper limit of $\sim 1$\% on the polarization of the lanthanide-free component. The low gamma-ray luminosity of GRB\,170817A\cite{gcn21507,gcn21509}, in spite of being located in the local Universe, seems to indicate that it is an off-axis event, although the possiblity of a peculiarly weak event cannot be ruled out. A natural prediction of off-axis scenario is that, following the deceleration of the outflow, the afterglow emission will be visible for the observer at very late times\cite{2014ARA&A..52...43B}. Though the off-axis afterglow emission could be linearly polarized\cite{2002ApJ...570L..61G}, during our polarimetric observations there is no evidence for such a component in the optical bands\cite{Pian}. The absence of polarization in our late-time optical data is therefore quite natural, while at earlier times the limits on the polarization of the lanthanide-free component are still within the allowed possibilities since the actual polarization fraction depends also on the degree of asymmetry of the outflows. Things could have been different in case of a NS/black-hole merging, since the ejecta are expected to be much more anisotropic than in the NS/NS case\cite{2015MNRAS.450.1777K,2015PhRvD..92d4028K}.

Our non-detection of linear polarization unambiguously due to a macronova emission up to very stringent limits is thus consistent with the theoretical expectations\cite{2015PhRvD..92d4028K} for this category of sources.
It also strengthens the identification of AT2017gfo with a macronova resulting from a NS/NS coalescence associated with a short GRB and a GW event and indirectly confirms that these sources are site of r-process production. If these events are  fairly common even in the local Universe\cite{2017arXiv170807008J}, it is likely that in the near future more macronovae will be observed enabling exploration of a variety of merging conditions and system parameters such as viewing angle, mass ratio, possible off-axis afterglow, etc. A spectro-polarimetric coverage that tracks the evolution of the phenomenon will shed light on possible deviations from the main expectations that are inaccessible with other techniques.

\begin{methods}

Our target, AT2017gfo, is the optical counterpart of the first GW signal detected by the Advanced LIGO and Virgo network\cite{LIGO2015,Acer2015} from the merger of a binary system of neutron stars, GW\,170817. Data were in general of excellent quality and were reduced following standard recipes, i.e. frames were bias-corrected and flat fielded and bad pixels flagged. Photometric analysis was carried out with standard aperture photometry. An additional complication arose since the target is located in the outskirts (at $\sim 10$\,arcsec) of the bright galaxy NGC\,4993 and for the latest epochs the galaxy light was comparable or more intense than the SSS17a brightness. We therefore modeled the external part of the galaxy with an analytical profile and effectively removed its contribution at the transient position. Photometry for the target was obtained by the acquisition frames calibrating with objects in the same field of view with the Pan-STARRS1 data archive\footnote{https://panstarrs.stsci.edu}.
Polarimetric observations were carried out by means of a Wollaston prism splitting the image of each object in the field into two orthogonal polarization components that appear in adjacent areas of the detector, a mask was used to avoid overlap of the two component images. For each position angle $\phi/2$ of the half-wave plate rotator, we obtain two simultaneous images of cross-polarization, at angles $\phi$ and $\phi+90^\circ$. We obtained observations at position angles $0^\circ, 22.5^\circ, 45^\circ$ and $67.5^\circ$ of the half-wave plate. This technique allows us to remove any differences between the two optical paths (ordinary and extraordinary ray) including effects of seeing and airmass changes. It is also possible to estimate the polarization introduced by galactic interstellar grains along the line of sight by studying the polarization of a large number of stars in the same detector area of the target to avoid a possible dependence of the instrumental polarization on the field of view position. The optical extinction in our Galaxy is $A_{\rm V} \sim 0.3$\,mag, and there is negligible extinction in the host galaxy\cite{Pian}. This implies that dust-induced polarization up to almost 1\% would be possible\cite{1975ApJ...196..261S}, although these are only statistical estimates and large variations on the main trend may be expected. The weighted average polarization shown by a set of stars nearby the transient turned out to be $P \sim 0.35$\% with a position angle $PA \sim 56^\circ$. Polarization is a positive quantity and at low signal to noise suffers from a statistical bias that was properly corrected\cite{2014MNRAS.439.4048P}. The last observation in the z-band did not allow us to derive constraining results because of the increasing difficulties due to the fading of the counterpart and the rapidly reducing visibility window. With the same setup we also observed polarized and unpolarized standard stars to convert position angles from the telescope to the celestial reference frame, and to correct for the small instrumental polarization introduced by the telescope. More details about imaging polarimetric data analysis can be found in, e.g.\cite{2012MNRAS.426....2W}.

The data that support the plots within this paper and other findings of this study are available from the corresponding author upon reasonable request. 
\end{methods}

%\bibliographystyle{naturemag}
%\bibliography{polrefs}

%% Here is the endmatter stuff: Supplementary Info, etc.
%% Use \item's to separate, default label is "Acknowledgements"

\begin{addendum}
 \item Based on observations collected at the European Organisation for Astronomical Research in the Southern Hemisphere 
under ESO programme 099.D-0116. We thank the ESO-Paranal staff for having carried out excellent observations under difficult conditions and in a hectic period.  We also acknowledge partial funding from ASI-INAF grant I/004/11/3. KW, ABH, RLCS, NRT acknowledge funding from STFC. JH was supported by a VILLUM FONDEN Investigator grant (project number 16599). YZF was supported by NSFC under grant 11525313. CGM acknowledges support from the UK Science and Technology Facilities Council. KT was supported by JSPS grant 15H05437 and by a JST Consortia grant. JM acknowledges the National Natural Science Foundation of China 11673062 and the Major Program of the Chinese Academy of Sciences (KJZD-EW-M06).
 \item[Author Contributions] All authors contributed to the work presented in this paper. S. Covino and K. Wiersema coordinated the data acquisition, analyzed the data and wrote the paper. A.B. Higgins, A. Melandri, P. D'Avanzo, E. Palazzi and N. Tanvir. contributed to the data-analysis. Y. Fan and K. Toma contributed to the theoretical discussion. C.M. Mundell contributed to the writing of the paper.
 \item[Competing Interests] The authors declare that they have no
competing financial interests.
 \item[Correspondence] Correspondence and requests for materials
should be addressed to S. Covino~(email: stefano.covino@brera.inaf.it).
\end{addendum}

%%
%% TABLES
%%
%% If there are any tables, put them here.
%%

\begin{table}
\centering
\caption{Results of the polarimetric campaign. Columns report the time after the GW event (17 August 2017, 12:41:04 UT\cite{gcn21505,gcn21509}), the Q/I and U/I Stokes parameters are corrected for the instrumental polarization, bias-corrected polarization\cite{2014MNRAS.439.4048P}, position angle and magnitudes obtained by the acquisition frames corrected for the Galactic extinction. Errors are at $1\sigma$ and upper limits are given at the 95\% confidence level.}
\medskip
\begin{tabular}{cccccc}
\hline
T-T$_{\rm GW}$ & Q/I & U/I & Polarization & Position Angle & Magnitude \\
\hline
(days) & & & (\%) & (deg) & (AB) \\
\hline
1.46 & $-0.0021 \pm 0.0008$ & $+0.0046 \pm 0.0007$ & $0.50 \pm 0.07$ & $57 \pm 4$ & $17.69 \pm 0.02$ \\
2.45 & $-0.0025 \pm 0.0016$ & $+0.0044 \pm 0.0032$ & $< 0.58$ & - & $18.77 \pm 0.04$ \\
3.47 & $-0.0009 \pm 0.0015$ & $+0.0034 \pm 0.0024$ & $< 0.46$ & - & $19.27 \pm 0.01$ \\
5.46 & $-0.0029 \pm 0.0033$ & $+0.0026 \pm 0.0050$ & $< 0.84$ & - & $20.39 \pm 0.03$ \\
9.48 & $+0.0412 \pm 0.0216$ & $-0.0095 \pm 0.0126$ & $< 4.2$ & -  & $20.69 \pm 0.11$ \\
\hline
\end{tabular}
\label{tab:pol}
\end{table}

\begin{table}
\centering
\caption{Observation log. Columns report the run number, the observation dates, the filter, the observed seeing, the airmass of the target, and Sun altitude below the horizon.}
\medskip
\begin{tabular}{ccccccc}
\hline
Run & Day & Exposure time & Filter & Seeing & Airmass & Sun Alt \\
\hline
& (August 2017, UT) & (s) & & (arcsec) & & (deg) \\
\hline
1 & 18.965-19.017 & 60 & R$_{\rm special}$ & 0.7-1.0 & 1.36-1.96 & 10.8 - 27.7 \\
2 & 19.967-19.996 & 90 & R$_{\rm special}$ & 1.5-2.0 & 1.39-1.68 & 11.4 - 20.9 \\
3 & 20.975-21.017 & 90 & R$_{\rm special}$ & 0.7-1.0 & 1.48-2.05 & 13.8 - 27.3 \\
4 & 22.973-23.018 & 120 & R$_{\rm special}$ & 0.7-1.0 & 1.51-2.20 & 13.0 - 27.7 \\
5 & 26.992-27.027 & 300 & z & 0.7-1.0 & 1.91-2.80 & 19.2 - 30.2 \\
\hline
\end{tabular}
\label{tab:obslog}
\end{table}

\begin{figure}
\includegraphics[width=18cm]{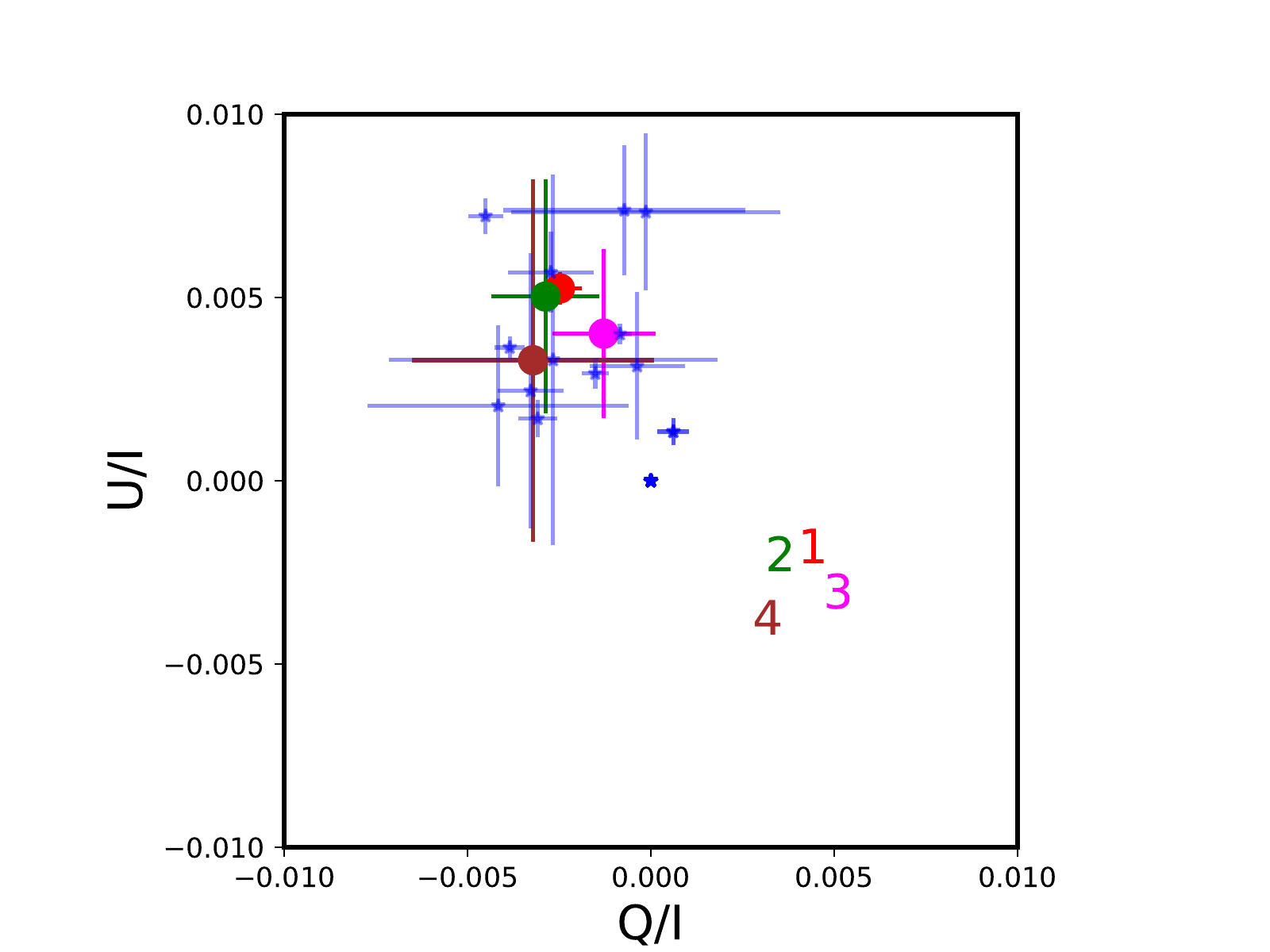} 
\caption{Q/U Stokes parameter plot for the optical transient and several field stars near to the transient. The reported numbers in the plot indicate the observation run as in Table\,\ref{tab:obslog}. The polarization of AT2017gfo (circles) is essentially indistinguishable from that shown by field stars (blue stars). Errors are at $1\sigma$. The Stokes parameters are a set of four parameters that describe the full polarization state of electromagnetic radiation. Q measures the difference between radiation intensity in the horizontal and the vertical direction of a given reference frame; U measures the difference between directions inclined by 45$^\circ$ and 135$^\circ$ with respect to the reference frame. I is the total intensity of the radiation. Together, Q and U therefore give the amplitude and angle of the linearly polarized component of the received intensity.} 
\label{fig:figpol}
\end{figure}

\end{document}